# A LIGHTWEIGHT TWO-LAYER BLOCKCHAIN MECHANISM FOR RELIABLE CROSSING-DOMAIN COMMUNICATION IN SMART CITIES

Xiangyu Xu and Jianfei Peng

College of Computer Science and Technology, Nanjing University of Aeronautics and Astronautics, Nanjing, China
xuxy@nuaa.edu.cn

## ABSTRACT

*The smart city is an emerging notion that is leveraging the Internet of Things (IoT) technique to achieve more comfortable, smart and controllable cities. The communications crossing domains between smart cities is indispensable to enhance collaborations. However, crossing-domain communications are more vulnerable since there are in different domains. Moreover, there are huge different devices with different computation capabilities, from sensors to the cloud servers. In this paper, we propose a lightweight two-layer blockchain mechanism for reliable crossing-domain communication in smart cities. Our mechanism provides a reliable communication mechanism for data sharing and communication between smart cities. We defined a two-layer blockchain structure for the communications inner and between smart cities to achieve reliable communications. We present a new block structure for the lightweight IoT devices. Moreover, we present a reputation-based multi-weight consensus protocol in order to achieve efficient communication while resistant to the nodes collusion attack for the proposed blockchain system. We also conduct a secure analysis to demonstrate the security of the proposed scheme. Finally, performance evaluation shows that our scheme is efficient and practical.*

## 1. INTRODUCTION

The development of Internet of Things (IoT) technology gives rise to many urban technologies, scenarios of which are more and more extensive: smart power grid [1, 2] relying on IoT infrastructure implemented intelligent monitoring and dispatching of power resources, simplifying complicated household electricity consumption procedures, and better coordinate supply and demand of urban electricity; With the help of IoT and edge computing, real-time systems such as Internet of vehicles [3], UAV [4, 5] have great potential value in intelligent transportation, intelligent logistics, urban security, agricultural monitoring and other aspects. The development and integration of these technologies enhance the bright future of smart city [6].

However, smart city faces many challenges [7-11]. First, smart city is a huge and complex system, and various devices are mainly not interoperable between them, which limits the collaborative work of the smart city. The above problem also exists in cross-city communication processes. Secondly, many data generated in smart city system are privacy-sensitive, such as user identity, purchase records, position information, etc. Therefore, it is challenging to implement a reliable, privacy guaranteed, heterogeneous cross-domain data communication network within and between smart cities. In addition, complex, frequent interaction systems tend to be fragile, which is threatened by plenty attacks, such as sybil attack allowing a malicious user or device to adopt multiple identities to occupy majority resources, Dos/DDoS attack which threats traditional centralized smart city network infrastructure, collusion attack which may control the entire smart city system, etc. These attacks may lead to the collapse of the entire smart city system or the loss of user's profit.

In order to solve these problems, there is a heated discussion in academia, producing many representative views: Rahman et al. proposed a IoT infrastructure based on blockchain and smart contract, which used to guarantee super wisdom city sustainable IoT security requirements in a shared economic [12]. Yu et al. proposed a distributed big data audit scheme based on lightweight blockchain to eliminate dependence of data audit on third-party authorities in traditional smart cities [13]. Sharma et al. proposed a smart urban automotive industry infrastructure based on the blockchain, which can be used to organize the manufacturing, supervision, maintenance and insurance services of self-driving vehicles, suitable for future smart contracts and smart applications [14]. These schemes prove that blockchain, as the infrastructure, can well organize the interactions within the distributed system and ensure the data integrity within the system. In addition, blockchain can credibly organize resource scheduling within a distributed system. These correspond to the dilemma faced by smart cities. Therefore, we could adopt blockchain to solve the problem of cross-domain communication within and between smart cities.

However, the application of blockchain in IoT has many drawbacks. First, as the price of high reliability, the cost of computation associated with traditional PoW based blockchain is so high that it doesn't apply to most IoT devices, because their computing resources are limited. The computing overhead of blockchain is mainly generated by the consensus protocol, but if we simply reduce the complexity of consensus computing, then the reliability of blockchain will decrease. How to balance the reliability and efficiency of blockchain could be a tricky issue. Second, blockchain is a public distributed ledger, and any participant has the right to access all information on the chain. This privacy protection brings challenges, so we need to implement an effective privacy protection mechanism on the basis of the blockchain infrastructure.

Our contribution: In this article, we designed a lightweight blockchain based on a two-layer blockchain to meet the cross-domain communication requirements of smart cities to address the previously mentioned issues of data security, privacy, and efficiency. Our contributions are as follows:

(1) We adopt a layered network structure to optimize the consumption of computing resources and realize access control. We use sub-chain and global-chain to organize the distributed architecture within the domain and between the domains/cities, so as to reduce the storage and computing pressure of devices.

(2) We designed a lightweight consensus protocol for the sub-chain, which enables the nodes to reach agreement on pooled data, dynamically manage the credit value of each node, and timely eliminate the Byzantine nodes. The improved PoS consensus protocol is adopted in the global-chain.

(3) According to the threat model we defined, we proved the security of the scheme.

Organization: The structure of this article are as follows. In section 2, we present some background that is related with our work. Our proposed methodology is illustrated in section 3. In section 4, some related work was presented. Finally, section 5 concludes this paper.

## 2. BACKGROUND

In this section, we introduce the basic concepts of smart cities and blockchain, and introduce threat models to help readers to understand our solution.

### 2.1. Smart City

#### 2.1.1. Preliminary

A smart city architecture could be simply abstracted into three levels: terminal equipment, edge devices, cloud server.

(1) Terminal equipment, such as various sensors, cameras and so on. These terminal devices directly collect data in the city, but their computing and storage resources are limited, so it is difficult to handle the computing tasks with high time complexity.

(2) Edge devices, such as roadside routers. These edge devices handle several terminal devices, which have a little more computing power, but still can't handle complex calculations. The calculation of edge equipment requires data collected by the terminal equipment in charge of its own or adjacent edge equipment.

(3) Cloud servers, all data are stored in the cloud server, and complex computing is also carried out in the cloud server.

A number of such smart city systems can be linked together to form smart city clusters. Cross-domain (which includes cross-city) communication is performed through cloud servers to disseminate data, while the edge and terminal equipment inside cities cannot directly interact with equipment outside cities. Here we divide a cloud server and the edge devices and IoT devices it manages into a domain. The construction of smart city cluster introduces cross-domain communication.

#### 2.1.2. Threat model

Due to the existence of malicious users or devices, as well as the unreliable communication channels, the complex network structure of smart cities often faces many security and privacy threats. Based on the smart city architecture and blockchain related preliminary introduced in the previous section, we define the threat models in this system as follows:

(1) Individual malicious device or user: by manipulating the malicious device, the attacker can release false information, lose packets, choose and forward packets and other attack means, thus damaging the normal operation of the system, or damage the reputation of other devices to gain more benefits in the reputation system.

(2) Device collusion attacks: in a smart city system, it may be easy to control edge or IoT devices, increasing the likelihood that an attacker will take control of as many devices as possible and make them work together for the same malicious purpose. It is not realistic to control more than half of the devices in the city, so we consider that attackers may interfere with the blockchain by controlling some devices for a period of time.

(3) DoS/DDoS attacks: considering the possible collusion of devices, we should also be aware that these conspired devices can perform DoS/DDoS attacks by sending large numbers of requests to the target device.

### 2.2. Blockchain

Blockchain [15] proposed by Satoshi Nakamoto in 2008 is generally regarded as a distributed, decentralized and highly trusted ledger. Blockchain enables accounting participants to reach consensus to transactions in a verifiable and secure way [16]. Its application scope is expanding from the financial field to the non-financial field. The blockchain infrastructure can be abstracted into three levels.

(1) Block: Participants listen to the broadcast of transactions, summarize them into the block and encapsulate it with hash, which ensures that the data integrity in the block can be verified.

(2) Blockchain ledger: Transactions are packaged into the block together with the hash of the previous block, which constitutes a chain structure that cannot be tampered with. Blocks are arranged in time sequence, and the order cannot be changed.

(3) Consensus agreement: In the blockchain network, each accounting node holds a copy of the blockchain, and in order to eliminate the influence of Byzantine nodes (on premise that the total number of nodes $n$ and the number of Byzantine nodes $f$ meet the requirement of $n > 2f + 1$), we need a consensus mechanism to make the accounting participants reach a consensus on the account book. Representative consensus protocols include Proof-of-Work (PoW) and Proof-of-Stake (PoS). PoW consensus protocol binds the workload of the block generation through hash collision calculation, and guarantees the credibility of the account book by assuming that it is difficult to master more than half of the computing resources of the whole network unilaterally. Obviously, PoW is not suitable for IoT devices. However, PoS determines the accounting right of nodes through the amount of equity held and the holding time, and guarantees the credibility of account books by assuming that it is difficult to unilaterally control more than half of the interests of the whole network. PoS avoids a lot of computing overhead and is suitable for IoT devices as well as edge devices.

## 3. METHODOLOGY

In this section, we will briefly introduce the scheme we designed, including distributed smart city system, intra-city communication based on sub-chain and cross-city communication based on global-chain, as well as the blockchain structure reconstructed based on smart city system.

### 3.1. Overview

(1) Sub-chain: The sub-chain is deployed inside the smart city, forming a peer-to-peer network between edge devices and edge devices, edge devices and cloud servers, and edge devices and sensors or actors. We deploy sub-chain on this distributed network. Sub-chain is a private blockchain used to track and record the interactions and reputation information of various nodes within a city, and has a series of policies to constitute internal and external access control management. As the city communication involves the privacy of many users, it is necessary to introduce the password mechanism to encrypt the data. Considering the limited storage resources of terminal devices such as sensors and referring to the Simple Payment Verification (SVP) node in the Bitcoin network, we only store the block header information in the terminal device, but store the complete copy of account book in the edge device. The edge device periodically uploads a copy of the ledger to the cloud server, releasing storage resources. Considering the limited computing resources of terminal and edge devices, we introduce a new consensus protocol based on average reputation value fusion to eliminate the influence of Byzantine nodes on the network.

(2) Global-chain: In order to facilitate the collaborative work between regions within cities and between smart cities, we consider the possibility of constructing smart city clusters. Cloud server/cloud server clusters in their respective cities can form a cross-domain (and cross-city) distributed network on which we try to deploy the parent chain. The parent chain adopts PoS consensus protocol.

### 3.2. Sub-chain: a reconstructed blockchain

Considering that the terminal and edge devices in the smart city system are resource-limited and low energy consumption, it is unreasonable to deploy the block chain based on PoW protocol on these devices. Therefore, we reconstructed a new blockchain for the sub-chain.

(1) Block structure: A block in sub-chain consists of header and body, whose structure is shown in Table 1, including the current block hash, the previous block hash, reputation root, strategy table, timestamp, and transaction root. The reputation tree is a tree data structure that uses the modified Merkle Patricia Trie structure to record the reputation value of nodes. Such data structure allows the node to record only the modified data in the block without affecting the computation of the reputation value, thus effectively reducing the memory footprint and computational complexity, as shown in the figure. The block body is composed of a reputation tree and a transaction tree, and the reputation value of each sub-chain node is recalculated after suspicious behavior, such as access violating the access control policy, and the creation of invalid blocks or transactions.

(2) Transaction structure: The transaction structure is shown in Table 2, where we define the microscale transaction data structure in detail, including the initiator device ID and target device ID, initiator signature, and additional data segments. This data structure takes up very little storage space, thus saving the limited storage resources in the sub-chain nodes. Since our reputation evaluation algorithm includes various weighting coefficients, we define a set of operations to describe transactions with different weighting coefficients. We classify transactions into the following five categories:

a) QUERY: the device queries specific information about the specified device by issuing a QUERY.

b) REPLY: the target device of the QUERY transaction takes advantage of the additional data segment of the REPLY transaction to REPLY.

c) UPDATE: the device broadcasts an UPDATE transaction to UPDATE the status of the device (such as a new action initiated by the actor, a new state detected by the sensor, a new device found by the edge device, etc.).

d) RATE: the equipment scores the reputation of a certain device through the reputation evaluation scheme, and announces the score by initiating the RATE.

e) ASSERT: a device broadcasts its own exception state by initiating the transaction.

Table 1. Composition of a block.

| **Contents** | **Size(bit)** | **Description** |
|---|---|---|
| CURRENT_HASH | 80 | Hash of current block |
| PRE_HASH | 80 | Hash of previous block |
| TMP | 24 | Timestamp |
| ROOT_REP | 80 | Root of reputation tree |
| ROOT_TRANS | 80 | Root of transaction tree |

Table 2. Composition of a block.

| **Contents** | **Size(bit)** | **Description** |
|---|---|---|
| TYPE | 4 | Type of transaction |
| ID_FROM | 8 | UID of sender service |
| ID_TARGET | 8 | UID of target device |
| SIG | 1024 | Signature of sender |
| ADD | 1024 | Note of transaction |

(3) Data processing: Every new node is assigned a pair of public and private keys before entering the domain. The unique ID of each node comes from its own public key to ensure the anonymity and non-repudiation of the framework. When a node receives a transaction, it must verify the signature of the transaction message to ensure the integrity and authorization of the transaction message, and transactions that are not validated are discarded. In order to minimize the computational overhead of terminal and edge devices, we use Keccak cite lightweight hash digest algorithm compared to the commonly used hash algorithm [17-19], because it is considered to have high performance in both program size and cycle counting. And we truncated the hash digest to 80 bits to save memory.

(4) Storage release mechanism: The maintenance of blockchain requires the continuous linking of new blocks to the end of the chain, which makes the ledger volume larger and larger. The capacity of terminals and edge equipment is limited, so we can infer the use frequency of equipment from data such as road conditions and people flow, and try to backup the data when the use frequency is low. After data backup is complete, terminal and edge devices need to free up memory in a timely manner.

## 3.3. Sub-chain: a consensus protocol based on reputation fusion

In this section, we introduce the credibility evaluation mechanism for sub-chain nodes and the consensus mechanism based on this mechanism.

### 3.3.1. Reputation evaluation mechanism

Our proposed trust evaluation mechanism maintains a credit score for each node in the sub-chain. When a new node joins a domain, other nodes in the network set an initial credit value of 100 for that node. When other nodes consider the node to be a Byzantine node (state exception, access and operation violating policy, etc.), the credit value is appropriately reduced according to the illegal operation of the node. This value can be increased when the node executes the correct command or feedback.

A node's credit is evaluated by other nodes that interact directly with it. Considering that different transactions have different characteristics, we add the weighted factor $W$ of the transactions into the credit evaluation calculation. We should also consider the timeliness of the data when evaluating the reputation of a node, so we specify that the node traverses and evaluates each transaction record at time *t*. At time *t*, the evaluation result of node *v* given by node *u* is $R_{u,v}(t)$, and we have:

$$R_{u,v}(t) = \sum_{i=1}^{C(u,v,t)} \sigma(t,i) \cdot Q(v,i) \cdot W(v,i) \;/\; \sum_{i=1}^{C(u,v,t)} W(v,i)$$

in which $C(u,v,t)$ represents the number of transactions generated between node *u* and node *v* before time *t*, $Q(v, i)$ represents the quality coefficient of *i*th transaction between *u* and *v*, $W(v, i)$ represents the significance coefficient of *i*th transaction between *u* and *v*, and $\sigma(t,i)$ represents the timelines coefficient of *i*th transactions. It is not difficult to understand that transactions with lower timeliness have less impact on the current system, so transaction timeliness is inversely proportional to its ability to affect the node's reputation. Let $t(i)$ be the moment when the *i*th transaction is completed, we define:

$$\sigma(t,i) = 1/(t(i) - t)$$

In this way, the credit value of each node is evaluated by other nodes that interact directly with it and updated dynamically. However, we need to consider that there is often more than one node to judge the reputation of node *v*, and that the evaluation of *v* by different nodes may vary according to the transactions with *v*. We believe that weighted reputation fusion is a good solution to this problem. Let $R(t_0)$ be the weighted credit score set of other nodes at time $t_0$, let $R_v(t)$ be the weighted credit score set of the newly calculated node *v* at time *t*, let $R_v$ be the credit score set of other nodes on node *v*. First, we removed the maximum and minimum values in order to reduce the impact of the maximum on the score without negotiation:

$$R_v^* = R_v \not\subset \{\max(R_v, \min(R_v))\}$$

After that, we take the reputation of other nodes as the weight, and calculate the weighted average of the reputation score of other nodes to node *v* as the current weighted reputation score of node *v*. We have:

$$R_v(\mathrm{t}) = \sum_{i=1}^{N} \frac{R_i(t_0)}{\sum_{j=1}^{N} R_j(t_0)} \cdot R_{i,v}(\mathrm{t})$$

in which $R_{i,v}(\mathrm{t})$ represents the reputation score of other node $i$ to node *v*. We take $R_v(\mathrm{t})$ the reputation of node *v* at time *t*.

### 3.3.2. Consensus protocol

In order to eliminate the influence of a small number of Byzantine nodes on the overall global ledger, so that all nodes reach a final agreement on the state of the ledger, we need a consensus agreement. In order to solve the contradiction between security and computing overhead mentioned above, we design a lightweight consensus protocol based on the credibility calculation method introduced above, which can avoid huge computing overhead while ensuring the anti-collusion performance of sub-chains.

Committee election: inspired by the DPoS protocol, we adopted the method of selecting a node delegation to be responsible for bookkeeping and generating new blocks, with the nodes in the delegation taking turns to be responsible for bookkeeping and block production in a specific order. The process of producing blocks does not require complicated mathematical calculations. We propose strategy 1 and strategy 2 for the sub-chain as the basis for delegation elections and mix the two strategies as the consensus agreement for the sub-chain.

a) Strategy 1: randomly select sub-chain nodes as miners. On the premise that more than half of the sub-chain nodes are trusted, we can randomly select a sub-chain node to be responsible for transaction collection, block packaging and release in the production block. Considering the limited computing resources of the terminal equipment, we can appropriately prefer the edge equipment node when selecting the miner.

b) Strategy 2: vote for sub-chain nodes as miners. Under the premise of a majority of the chain, we can select first *n* nodes for candidate nodes according to descent ranking of nodes' reputation, constitute a set *S*, and choose *K* online nodes to constitute the final mining executive set *E*. Nodes in *E* take turns as the miner, until all nodes in *E* were travelled, and reselection of *E* should be triggered. If the ledger is bifurcated, then we can assume that the credit value of nodes within *S* will change, so we need to re-select *S*.

The advantage of strategy 1 is that it is difficult to predict the miners' nodes in each round of production block, so the influence of Byzantine nodes in the sub-chain on the whole system is minimized. The advantage of strategy 2 is that there will not be frequent and complex miner election behavior for a period of time, thus achieving optimal sub-chain performance. Strategy combining strategy 1 and strategy 2, we finally designed the consensus protocol as shown in Algorithm 1.

In this way, users can adjust each threshold to adapt to different security needs or security environment, and the sub-chain will adjust the system state timely according to the security threshold set by users, and actively isolate the Byzantine nodes with greater influence, to further reduce the influence of Byzantine nodes.

**Algorithm 1: Consensus Protocol**

Input:
  Counting Threshold $T_1$;
  Duration Threshold $T_2$, $T_3$;
  Reputation Threshold $T_4$;
  $S$ Election Counter $\beta_1$;
  Timer $\beta_2$;
Output:
  Global Consensus
  begin
  start $\beta_2$
  for $\beta_1$ = 0 to $T_1$ do
    exec Strategy 2
  end for
  stop $\beta_2$
  if $\beta_1 > T_1$ and $\beta_2 > T_2$ then
    for $v$ in whole network do
      $R_v(\mathrm{t}) \leftarrow R_v(\mathrm{t})/2$
    end for
    clear $\beta_2$ and start
    exec Strategy 1
    if $\beta_2 > T_3$ then
      stop $\beta_2$
      isolate $v$ in whole network which $R_v(\mathrm{t}) < T_4$
    end if
  end if
  clear $\beta_1$, $\beta_2$
  goto 2
  end

### 3.3.3. Block Release

Another important issue for leveraging the blockchain technique in the resource constrained IoT network is storage release problem. As in the blockchain technique, each note will store a copy of the whole ledge, the size of the blockchain system will grow with the block increase. However, in the IoT systems, each IoT devices are with limited storage. Thus, we need a mechanism to release the past block in the IoT devices without decreasing the security of the blockchain system. In our design, we leverage the constant release method to address this problem. With a constant time, all nodes will agree with the current status of the blockchain systems. All nodes will store a consensus block that stores the hash value of the previous blocks, and this consensus block will act as a new genesis block. The block data will be stored in a server for inspection when it is needed in later.

# 4. SECURITY ANALYSIS AND PERFORMANCE EVALUATION

## 4.1. Security Analysis

According to our design of lightweight consensus scheme, if the sub-chain in the frequency range of users receive the books of bifurcate, then substring automatically into a safer consensus strategy, and isolate the credibility not timely recovery of nodes (we can assume that the nodes have has great influence on the system). Therefore, if a Byzantine node wants to survive in a sub-chain for a long time, it must compress the length of its abnormal state to a range acceptable to the user.

From the reputation evaluation mechanism, it can be seen that the higher the credibility of the node, the greater the weight of other nodes. Obviously, under the premise that the number of Byzantine nodes is less than 1/2 of the number of summary points, the colluding Byzantine nodes cannot play a decisive role in the credibility of the node. To sum up, the influence of Byzantine nodes on system security is likely to be tolerated by users or external systems.

## 4.2. Malicious Server

As mentioned before, a malicious server might damage the availability of the whole system in two methods. Broadcasting fake information which might cause usage accidents could be defended by the novel reputation evaluation scheme. It mainly because the activities of each device in the network are being evaluated to build a trust rating scheme and the receiver accepts or drops the message according to the reputation value of the server. Thus, those fake information and unfair reputation report messages could be blocked with high probability.

## 4.3. Dos/DDos Attacks

In the design of our system architecture, the Dos/DDos attack can be mitigated by evaluating the reputation value of each node in the blockchain. If a malicious node executes the Dos/DDos attack, it is needed to generate amount of blocks to consume other nodes' resource. However, when a new block is broadcasted to the blockchain network, it will consume a predefined amount of reputation of the block holder. When the block holder's reputation decreased to an low bound, all its blocks broadcasted to the blockchain network will be discarded by other nodes. In such a manner, each node in the blockchain network can only issue a limited number of blocks, and thus the Dos/DDos attack is mitigated by our design.

## 4.4. Performance Evaluation

In this subsection, we conduct an experiment to evaluation the performance of our proposed blockchain system. We implemented the nodes' reputation value in three different manners: all nodes are with the same reputation value and this value is constant, a random reputation is set for each node in the system, and our presented reputation computation method.

As shown in Figure. 1, with the time increase, the reputation of a malicious will go down in our method and the random reputation method. Moreover, the reputation of a malicious in our method is lower than that of the other method. Thus, the attack executed by the malicious node can be protected better in our design.

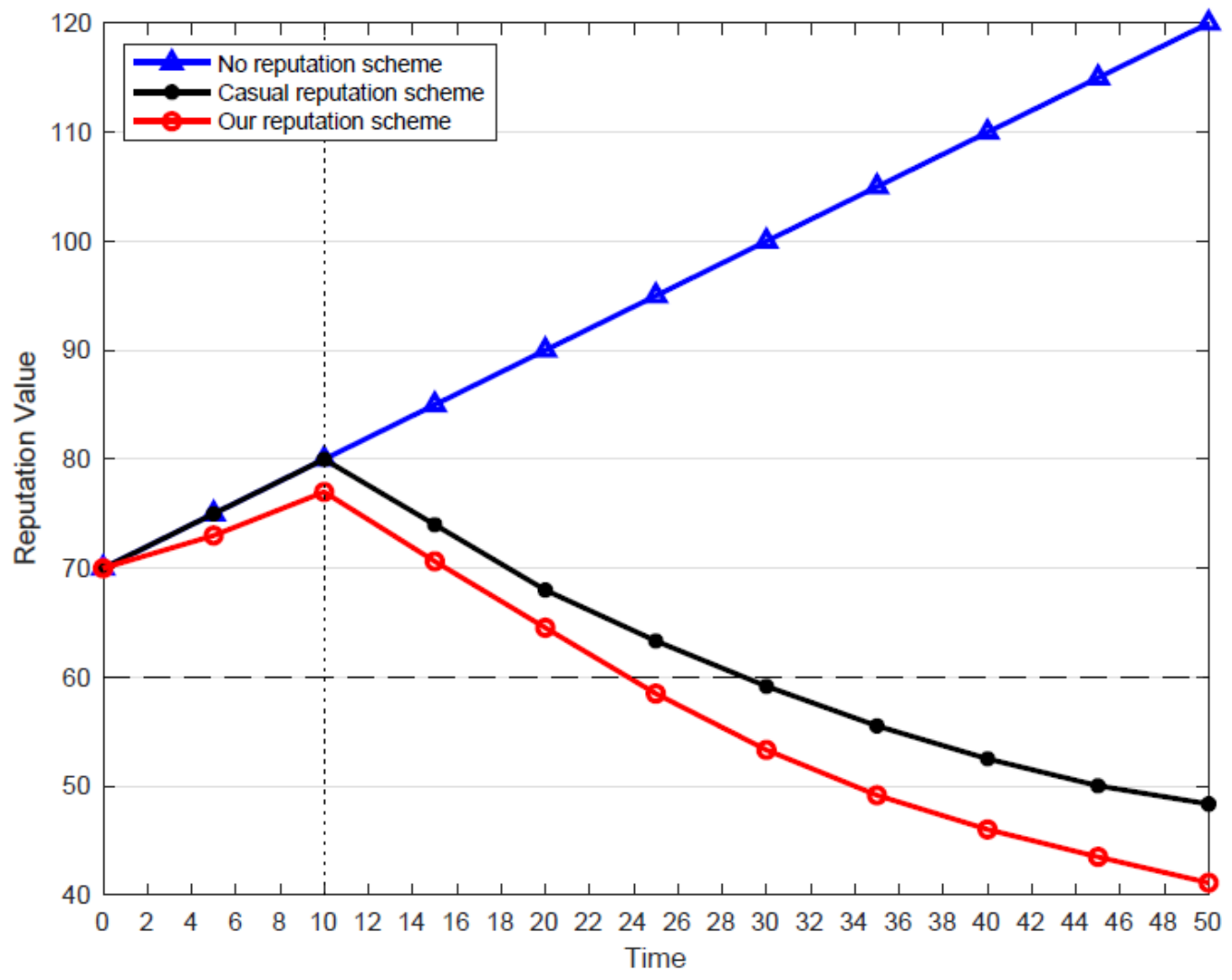

Figure 1. Reputation fluctuation of a malicious node

## 5. Related work

For the security and privacy protection of smart cities, the academia has proposed many feasible schemes. Mehdi Gheisari et al. adopted the smart city IoT system based on software custom network (SDN) and proposed an effective privacy protection method [20]. Yin et al. proposed an attack vector assessment model based on vulnerability, path and action, and proposed a formal representation and quantitative assessment method for network security risk assessment of critical information infrastructure in smart cities [21]. Wang et al. ensured information security at the hardware level [22]. Badii et al. proposed an Internet of things and smart city platform that conforms to the GDPR specification, and claimed to have verified the reliability of the scheme in practical application [23]. Some researchers also proposed to use ontology to define an edge computing network to protect privacy [24]. Khatoon et al. proposed an efficient, secure, bilinear pair-based, unlinked, and mutually authenticated key protocol to ensure the privacy security of telecare medicine information system (TMIS) [25]. Duan et al. focused on modelling privacy content from multiple sources, mapping them to the data, information, and knowledge types of resources in the well-known DIKW architecture [26]. Sucasas et al. proposed a protocol based on OAuth 2.0 for privacy protection in smart city mobile applications [27]. Gope et al. proposed a lightweight, private-protected RFID authentication scheme for distributed Internet of things infrastructure based on secure localization services for smart urban environments [28].

On the other hand, the blockchain technique has been leveraged in many other IoT application scenarios due to the development of embedded processors, high speed network technique and artificial intelligence. Adopting the blockchain technique in the Internet of Vehicles (IoV) is an emergence representative. The first decentralized trust management mechanism based on blockchain for IoV which leveraged the proof ow stake and proof of work consensus protocol was introduced by Yang et al. [29]. In their scheme, the trust value model was used to evaluate a node pow to write a new block. Ma et al. [30] introduced a novel lightweight blockchain system for the IoV system. In their design, the reputation of a node is viewed as its stake. To ensure real time communication, they designed two separated chains to deal with the outside communication and inside communication for a vehicle. Later, many different blockchain platforms was designed to realize secure, robust and privacy communications in the IoV network.

## 6. Conclusion

In this paper, we investigate the probability of blockchain infrastructure organizing a cross-domain communication in smart city, proving that blockchain is of great future to solve privacy preserving and data integrity issues in smart city. To solve contradiction between computational-sensitivity of blockchain and limited computational resources in smart city, we proposed two-layer blockchain architecture: sub-chain and global-chain. To fit IoT devices, we reconstructed the data structure of blockchain and proposed a reputation-based consensus scheme inspired by DPoS consensus protocol. Security analysis and related experiment are performed to prove the effectiveness of our scheme. We consider our scheme as a new thought that can solve data security issues in smart city.

**Authors**

Xiangyu Xu
received his Bachelor's degree in 2017, from Nanjing Tech University, China. He is currently a master student in College of Computer Science and Technology, Nanjing University of Aeronautics and Astronautics, China. His research interests include system security and IoT security.

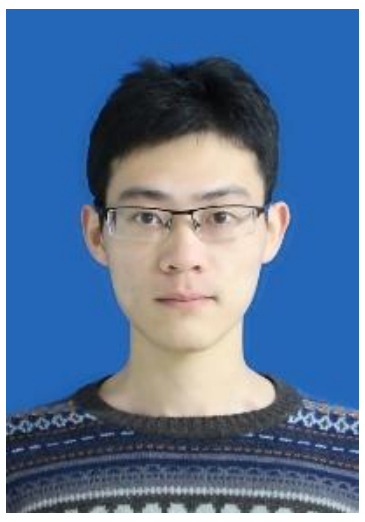

Jianfei Peng
received his Bachelor's degree in 2018, from Nanjing University of Aeronautics and Astronautics, China. He is currently a master student in College of Computer Science and Technology, Nanjing University of Aeronautics and Astronautics, China. His research interests include Cloud Security, System Security and IoT Security.

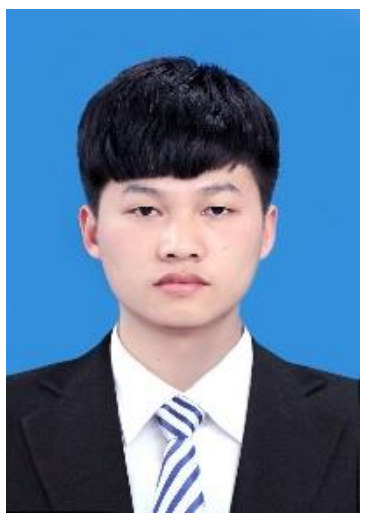